\documentclass[aps,prstab,reprint,superscriptaddress,showpacs]{revtex4-1}

\usepackage{graphicx}
\usepackage{multirow}
\usepackage{romannum}

\begin{document}
\pagenumbering{arabic}

\title{High Bandwidth Pickup Design for Bunch Arrival-time Monitors \\ for Free-Electron Laser}
\author{Aleksandar Angelovski}
\email[]{angelovski@imp.tu-darmstadt.de}
\affiliation{Technische Universit\"at Darmstadt, Institute for Microwave Engineering and  Photonics, \\ Merckstrasse 25, 64283 Darmstadt, Germany }

\author{Alexander Kuhl}
\email[]{kuhl@gsc.tu-darmstadt.de}
\affiliation{Technische Universit\"at Darmstadt, Graduate School Computational Engineering, \\ Dolivostrasse 15, 64293 Darmstadt, Germany}
\affiliation{Deutsches Elektronen-Synchrotron DESY, Notkestrasse 85, 22607 Hamburg, Germany}

\author{Matthias Hansli}
\affiliation{Technische Universit\"at Darmstadt, Institute for Microwave Engineering and Photonics, \\ Merckstrasse 25, 64283 Darmstadt, Germany }
\author{Andreas Penirschke}
\affiliation{Technische Universit\"at Darmstadt, Institute for Microwave Engineering and Photonics, \\ Merckstrasse 25, 64283 Darmstadt, Germany }
\author{Sascha M. Schnepp}
\affiliation{Technische Universit\"at Darmstadt, Graduate School Computational Engineering, \\ Dolivostrasse 15, 64293 Darmstadt, Germany}
\author{Michael Bousonville}
\affiliation{Deutsches Elektronen-Synchrotron DESY, Notkestrasse 85, 22607 Hamburg, Germany}
\author{Holger Schlarb}
\affiliation{Deutsches Elektronen-Synchrotron DESY, Notkestrasse 85, 22607 Hamburg, Germany}
\author{Marie Kristin Bock}
\affiliation{Deutsches Elektronen-Synchrotron DESY, Notkestrasse 85, 22607 Hamburg, Germany}
\author{Thomas Weiland}
\affiliation{Technische Universit\"at Darmstadt, Institut f\"ur Theorie Elektromagnetischer Felder, TEMF, \\Schlossgartenstrasse 8, 64289 Darmstadt, Germany }
\author{Rolf Jakoby}
\affiliation{Technische Universit\"at Darmstadt, Institute for Microwave Engineering and Photonics, \\ Merckstrasse 25, 64283 Darmstadt, Germany }



\date{\today}

\begin{abstract}
  
In this paper, we present the design and realization of high bandwidth pickup electrodes with a cutoff frequency above $40$~GHz. The proposed cone-shaped pickups are part of a bunch arrival-time monitor (BAM) designed for high ($>500$~pC) and low ($20$~pC) bunch charge operation mode providing for a time resolution of less than $10$~fs for both operation modes. The proposed design has a fast voltage response, low ringing, and a resonance-free spectrum. For assessing the influence of manufacturing tolerances on the performance of the pickups, an extensive tolerance study has been performed via numerical simulations. A non-hermetic model of the pickups was built for measurement and validation purposes. The measurement and simulation results are in good agreement and demonstrate the capability of the proposed pickup system to meet the given specifications.

\end{abstract}


\maketitle


\section{Introduction}

High gain free-electron lasers (FELs) are able of generating ultra short X-ray pulses with a duration in the femtosecond range \cite{Ackermann}. In order to provide an optimal operation of the FEL for pump-probe experiments or for seeding using external laser systems, the arrival-time of the bunches has to be synchronized with femtosecond precision. An accurate measurement of the bunch arrival-time, thus, is essential for the synchronization process.
 
Currently, there are few bunch arrival-time monitors (BAMs) available, which achieve or have the potential to achieve sub-$10$~fs time resolution. An arrival-time monitor using a cavity as a beam pickup and a radio frequency (rf) based phase detection is described in \cite{Anderson}. Such a detection scheme is realized at the LCLS free-electron laser at SLAC. It allows for a sub-50 femtosecond synchronization between a laser and the X-rays for pump-probe experiments \cite{Brachmann}.
Another type of arrival-time monitor featuring an electro-optical crystal inside the beam pipe for measuring the cross-correlation of the coherent terahertz radiation generated by the undulator and an external laser pulse is presented in \cite{Tavella}. This type of monitor can be applied only at the end of an undulator.

At the free-electron laser at DESY, Hamburg (FLASH), the installed BAMs have an intrinsic time resolution better than 10 fs for bunch charges above $500$~pC \cite{Loehl}. These arrival-time monitors combine a transient beam induced pickup signal with an electro-optical signal detection scheme as proposed in \cite{Loehl1}. In this scheme the achievable time resolution is proportional to the steepness of the output voltage at the first zero-crossing. The steepness in turn scales with the bunch charge leading to significant performance degradations for low charge FEL operation. Details are presented in \cite{bock}.

With the extension of FLASH~\Romannum{2} and the European X-Ray Free Electron Laser Project (XFEL), a low charge operation mode with bunch charge of $20$~pC is planned. In order to satisfy the sub-$10$~femtosecond resolution demands for high and low charge operation, the current BAM design needs to be upgraded. 

In this paper we present the design and the realization of high bandwidth cone-shaped pickup electrodes as a part of the BAM. The proposed pickup electrodes enable high resolution arrival-time measurements for the low bunch charge as well as for high bunch charge mode of operation. 

The remainder of this paper is organized as follows. The design procedure and simulation results along with a performance characterization of the pickup are given in Sec.~\ref{Design}. In Sec.~\ref{Tolerance}, results of a tolerance study are presented, which was conducted for assessing the influence of fabrication tolerances on the performance of the pickup.
A non-hermetic prototype was built for proving the concept. Measurements and comparison to simulation results are shown in Sec.~\Romannum{4}. The integration of the pickup in the BAM and the cabling procedure is proposed in Sec.~\ref{Cabling}.

\section{Pickup Design and Simulation}
\label{Design}

The requirements on the pickup, which need to be fulfilled for low and high charge operation are given in Table~\ref{tab:Param}. An output voltage slope at the first zero-crossing of more than $300$~mV/ps is required. In frequency domain this corresponds to a cutoff frequency of 40 GHz. The ringing of the voltage signal is defined as

\begin{equation}
R_{T_0} := \frac{\max\left (| U(t) | \right ) | _{T_0 \leq t \leq T}} {\frac{1}{2} V_{\mathrm{pp}}}
\label{eqn:definition_ringing}
\end{equation}
where $T_0$ is a freely chosen time offset with respect to the zero crossing, $V_{pp}$ is the peak-to-peak voltage, and $T$ is the simulation time.
This definition of the ringing is adapted from the design specifications.
These require the ringing to be less than $0.01$\% of the peak amplitude after $222$~ns, which corresponds to the minimum bunch spacing for the European XFEL.
\begin{table}[h]
\caption{\label{tab:Param}Design parameters and requirements.}
\begin{ruledtabular}
\begin{tabular}{ll}
  Parameter & Value \\
\hline
  Output voltage slope & $>300$~mV/ps\\
  Bandwidth & $>40$~GHz\\
  Ringing after $222$~ns & $<0.01$\%~of $V_{pp}$\\
  Bunch  charge & $20$~pC - $1$~nC
\end{tabular}
\end{ruledtabular}
\end{table}

The current pickup design introduced in \cite{Kirsten} does not fulfill the requirements for high and low charge operation mode. Namely, the simulation of the current pickup shows a voltage slope of $69.7$~mV/ps and strong ringing.
\subsection{Design}

The rf properties of the pickup are defined by its shape, the feedthrough material, the connectors, and the cables. However, the pickup itself has the largest influence on the performance of the system.  

\begin{figure}[h]
 \centering
	\includegraphics[width=8.6cm]{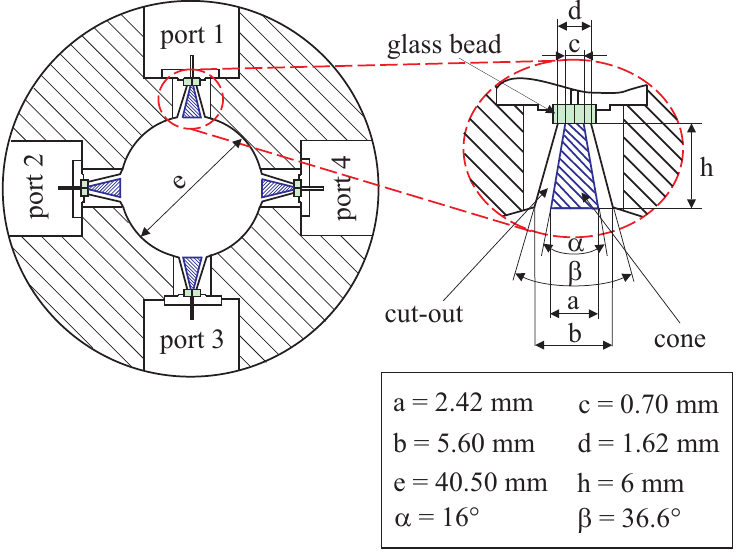}%
 \caption{\label{Cross}Cross section of the cone-shaped pickup with the dimensions.}
 \end{figure}

We propose a tapered coaxial structure, which comprises a cone-shaped pickup electrode with the corresponding cut-out, as shown in Fig.~\ref{Cross}. Unlike the classical button type pickup, the cone-shaped pickup avoids resonances within the pickup due to the tapered transition from the beam pipe to the connector.
The diameter of the cut-out is denoted by $b$, and the tip diameter of the cone-shaped pickup is denoted by $a$.    

The optimal ratio of $b/a$ is $2.3$ as it provides for $50~\Omega$ impedance in vacuum. This ratio of the cone to cut-out diameter is maintained at every cross-section from the connector to the pickup tip. It is, thus, matched to the impedance of the system and provides for a resonance-free spectrum up to $40$~GHz. For any other value of $b/a$ reflections occur, leading to ringing of the output voltage signal with a higher amplitude and larger decay time. 
The feedthroughs used in this design have a dielectric with a relative permittivity of $\varepsilon_{r}\sim4.1$. These feedthroughs are designed for $2.92$~mm  connectors (K-connectors) specified for operation up to $40$~GHz.

For the design and simulation of the cone shaped pickup (Fig.~\ref{Cross}) we used the CST PARTICLE STUDIO\textregistered\ software package, which allows for computing the electromagnetic field of a particle beam in time domain.
As one feature, the software allows to compute the time dependent voltage along a user defined path.
We defined such a path at the pickup connector, specifically from the inner to the outer conductor and extracted the pickup output voltage. In order to determine the voltage slope at the first zero crossing and the amplitude of the ringing, an automatic analysis was developed and implemented as a post-processing step by means of Visual Basic for Applications (VBA) macro scripts.
In the following, all simulations are carried out using Gaussian bunches with a longitudinal standard deviation of $\sigma$~=~1~mm and 20~pC bunch charge.
\subsection{Convergence study}
\begin{figure}[h]
   \centering
   \includegraphics*[width=86mm]{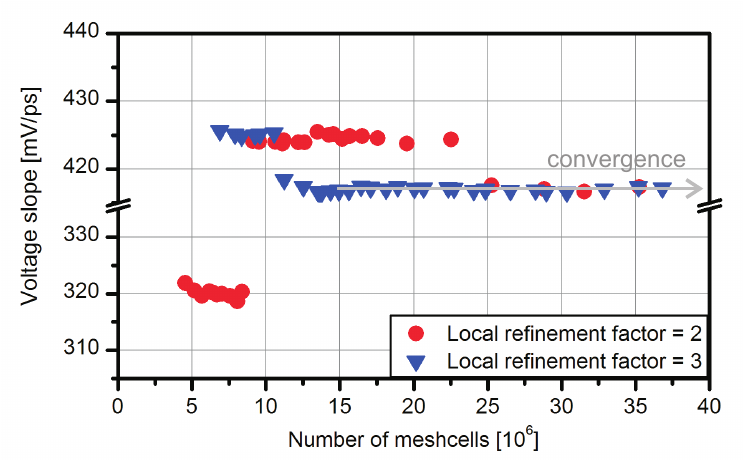}
   \caption{Plot of the slope vs.~the number of mesh cells and two settings of the local mesh refinement. Two symmetry planes are used for the simulation such that only one quarter of the model has to be considered. The slope converges to approximately $417$~mV/ps.}
   \label{fig:konvergenzanalyse_slope1}
\end{figure}

In order to ensure correct simulation results a convergence study was conducted.
The design was simulated using a series of refined computational meshes until a steady state of the investigated quantities was obtained.
As one example of the quantities considered in this study, we present the convergence of the output voltage slope at the first zero crossing to a steady state of approximately $417$~mV/ps under mesh refinement (cf.~Fig.~\ref{fig:konvergenzanalyse_slope1}).

The slope is a highly sensitive quantity because its value depends on the derivative of the voltage signal.
For obtaining refined meshes, global as well as an additional local refinement was applied.
The region around the cone and the glass bead feature very small dimensions in comparison to the beam pipe.
Accurate results, critically hinge on a good mesh resolution around these parts.
Therefore, they were embedded in a local mesh refinement using refinement factors of two and three corresponding to a division of the respective grid into 8 and 27 cells respectively.
Furthermore, a local refinement by a factor of four was tested, but in this case the number of time steps and, thus, the computing time increases drastically. For all refinement methods the same steady state value is obtained.
However, large deviations for insufficient mesh resolutions can be observed.
The results of the convergence study are shown in Fig.~\ref{fig:konvergenzanalyse_slope1}.
It is readily seen that using a local refinement factor of three the slope reaches a steady state if at least $13$ million mesh cells are used. Using a refinement factor of two on the other hand requires at least $24$ million mesh cells for obtaining steady state.
Based on the convergence study we chose the mesh resolution such that a signal of $40$~GHz is sampled by approximately $32$ mesh lines per wavelength and applied a local refinement factor of three in the vicinity of the cone and the glass bead.
These settings were used consistently throughout all following simulations. The simulations were performed using two symmetry planes such that only one quarter of the model has to be considered. By increasing the number of mesh cells (from left to right) the computed value of the slope converged to approximately $417$~mV/ps. For insufficient mesh resolutions large deviations can be observed. These are attributed to a small number of cells at the feedthrough, which are automatically filled up with metal. 

\subsection{Simulation results}

\begin{figure}[h]
   \centering
	\includegraphics[width=86mm]{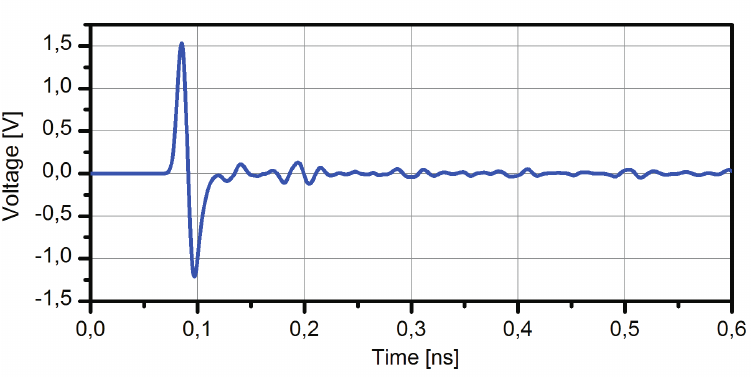}
	\includegraphics[width=86mm]{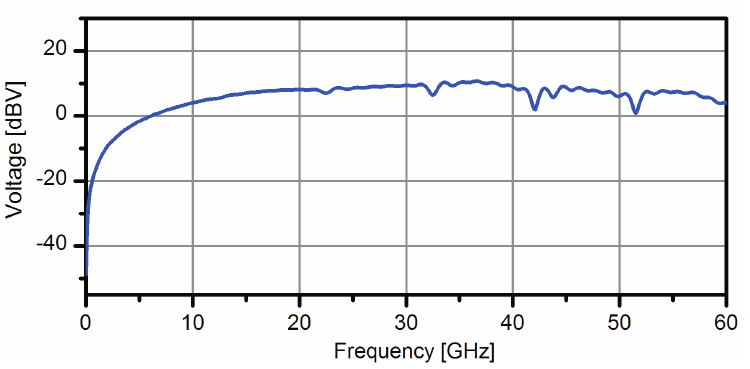}

   \caption{Simulation results of one pickup in time domain (top) and the respective frequency spectrum normalized by the spectrum of the particle beam (bottom).}
   \label{fig:zeitverlauf_und_spektrum}
\end{figure}

The result of the simulation of the proposed design in time domain and the corresponding frequency spectrum is shown in Fig.~\ref{fig:zeitverlauf_und_spektrum}. The actual lossy material parameters were employed in the simulations. Namely, the cone was designed using the parameters from covar, the flanges from stainless steel and the glass bead from Corning 7070 glass. 

The time domain voltage signal shown in Fig.~\ref{fig:zeitverlauf_und_spektrum}~(top) has a voltage slope of $416.78$~mV/ps at the first zero crossing. The result exceeds the requirements for the voltage slope given in Tab.~\ref{tab:Param}. In comparison to the current pickup \cite{Kirsten} the slope was increased by factor of six while ringing was drastically reduced at the same time. The time distance between the two peaks of the voltage signal, which determines the dynamic range of the monitor is approximately $13$~ps. The amplitude of the ringing is less than $1$\% from the peak amplitude after $0.6$~ns. 
Typically, the simulations cover a time span of $0.6$~ns using $T_0 = 0.3$~ns in (\ref{eqn:definition_ringing}) for characterizing the ringing. There are several factors which influence the ringing. The resonances within the pickup, the cross-talk, and the interaction of the beam with the surrounding environment. The first two can be subject to optimization, and they have the largest influence. The last one cannot be avoided but the contribution to the ringing is significantly lower compared to the first two. 

The spectrum of the voltage signal [see Fig.~\ref{fig:zeitverlauf_und_spektrum}~(bottom)] has an approximately flat characteristics up to $40$~GHz. The small kinks in the spectrum around $23$~GHz, $33$~GHz, and $43$~GHz are due to the cross-talk between the pickups, which can be observed around $0.2$~ns and $0.3$~ns in the time domain signal [see Fig.~\ref{fig:zeitverlauf_und_spektrum}~(top)]. The designed cone-shape pickup has a resonance-free spectrum and provides for an output voltage signal with a low ringing amplitude and a fast decay.

\section{Tolerance study}
\label{Tolerance}
Due to manufacturing tolerances the actual dimensions of the pickup might differ from the design values.
The tolerance study reveals the sensitivity of the output voltage signal with respect to several manufacturing tolerances.
A list of some of the varied parameters and the variation range is shown in Table~\ref{tab:toleranzanalyse}.
\begin{table}[h]
   \caption{\label{tab:toleranzanalyse}List of the design parameters and the variation ranges as investigated in the tolerance study.}
\begin{ruledtabular}
   \begin{tabular}{lll}
Parameter & Value & Range of variation \\
\hline
cutout small radius				&$0.8125$\,mm		&$\pm$ 0.2\,mm\\
cone small radius				&$0.35$\,mm		&$\pm$ 0.15\,mm\\
cone length					&$6.00$\,mm		&$\pm$ 0.2\,mm\\
cutout angle					&$36.6^\circ$		&$\pm$ 2.5$^\circ$\\
cone angle					&$16.0^\circ$		&$\pm$ 2.5$^\circ$\\
$\epsilon_r$ of the glass			&$4.15$			&$\pm$ 0.2
   \end{tabular}
\end{ruledtabular}
\end{table}
Based on the results of this study the production tolerances can be specified. We investigated the influence of manufacturing tolerances onto the slope of the output signal at the first zero crossing and the ringing. 

\begin{figure*}[tb]
   \centering
   \includegraphics[width=1\textwidth]{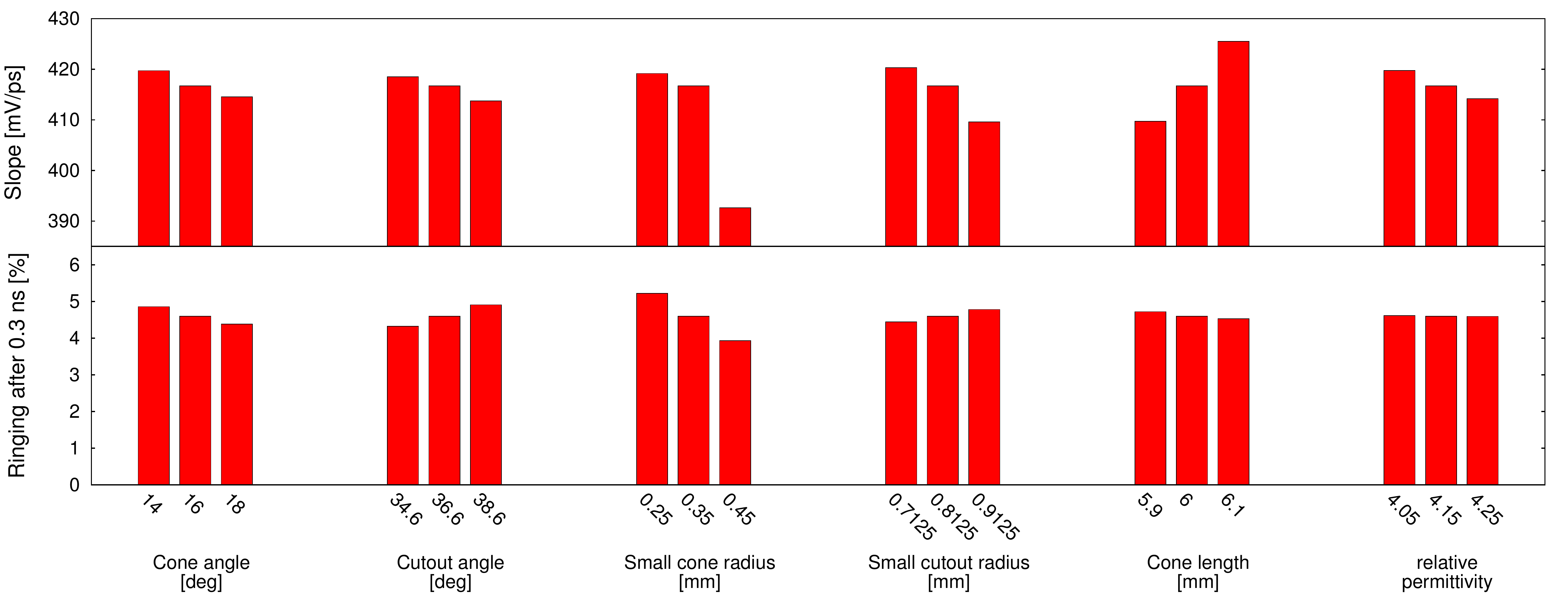}
   \caption{Influence of geometric variations on the slope (top) and on the ringing (bottom).}
   \label{fig:toleranzanalyse_slope_und_ringing}
\end{figure*}

The results of the tolerance study are shown in Fig.~\ref{fig:toleranzanalyse_slope_und_ringing}.
They indicate that a higher slope can be obtained by, e.g., increasing the cone length or changing the cone angle.
Since in the former case the cone reaches into the beam pipe and in the latter case the matching to $50~\Omega$ is lost, none of these modifications are permissible. 
\begin{table}[h]
   \caption{\label{tab:einfluss_mit_fit_funktionen_ermittelt}Computed change of the slope with quadratic fit based on the tolerance study results.}
\begin{ruledtabular}
\begin{tabular}{l|cc}
Parameter & Deviation & $\Delta$ Slope [mV/ps] \\
\hline
\multirow{4}{*}{\parbox{2cm}{cone angle \newline $\alpha$}} 
& -1.0$^\circ$	&+1.35\\
& -0.5$^\circ$	&+0.69\\
& +0.5$^\circ$	&-0.74\\
& +1.0$^\circ$	&-1.52\\
\hline
\multirow{4}{*}{\parbox{2cm}{cutout angle \newline $\beta$}}
& -1.0$^\circ$	&+1.18\\
& -0.5$^\circ$	&+0.62\\
& +0.5$^\circ$	&-0.67\\
& +1.0$^\circ$	&-1.38\\
\hline
\multirow{4}{*}{\parbox{2cm}{small cone radius \newline c}}
& -50$\mu$m	&+4.00\\
& -20$\mu$m	&+2.26\\
& +20$\mu$m	&-3.13\\
& +50$\mu$m	&-9.46\\
\hline
\multirow{4}{*}{\parbox{2cm}{small cutout radius \newline d}}
& -50$\mu$m	&+1.22\\
& -20$\mu$m	&+0.61\\
& +20$\mu$m	&-0.78\\
& +50$\mu$m	&-2.26\\
\hline
\multirow{4}{*}{\parbox{2cm}{cone length \newline h}}
& -50$\mu$m	&-4.06\\
& -20$\mu$m	&-1.63\\
& +20$\mu$m	&+1.63\\
& +50$\mu$m	&+4.08\\
\hline
\multirow{4}{*}{$\epsilon_r$ of the glass}
& -0.10		&+2.88\\
& -0.05		&+1.42\\
& +0.05		&-1.38\\
& +0.10		&-2.72
\end{tabular}
\end{ruledtabular}
\end{table}

We fitted the results of the tolerance study for each geometric parameter, e.g. the length of the cone, to a quadratic function.
Having the fitting functions at hand, the influence of each geometry parameter onto the slope as well as
all other quantities of interest can be determined for any parameter deviation within the investigated range of variation (cf.~Table~\ref{tab:toleranzanalyse}).
In Table~\ref{tab:einfluss_mit_fit_funktionen_ermittelt} the changes of the output signal slope of the most sensitive parameters are summarized assuming typical manufacturing tolerances.
The most sensitive parameter regarding the voltage signal slope is the small cone radius $c$ (see Fig.~\ref{Cross}), which should be produced with a precision of $\pm$20~$\mu$m.
For all other parameters a tolerance of $\pm$50~$\mu$m is acceptable. Regarding the opening angle of the cone and cutout a tolerance of $1^\circ$ should be achieved.
Based on this step of the tolerance study, we performed a worst case scenario simulation. In this simulation tolerances are allowed not only for one parameter at a time but for all parameters at the same time. We set tolerances of $\pm$20\,$\mu$m for the small cone radius and $\pm$50\,$\mu$m respectively $1^\circ$ angle deviation from the design values.
The results of the worst case simulation provide a slope of $404.2$~mV/ps, which corresponds to a decrease of $3.02$\%.
The most sensitive parameter concerning the ringing (see $R_{0.3 ns}$ in Fig.~\ref{fig:toleranzanalyse_slope_und_ringing}) is also the small cone radius.
However, long term simulations show that the decay of the ringing is well inside the specifications.
Ringing is thus no primary concern.

\section{Non-hermetic pickup prototype}
\label{Prototype}

\begin{figure}[h]
 \centering
 \includegraphics[width=86mm]{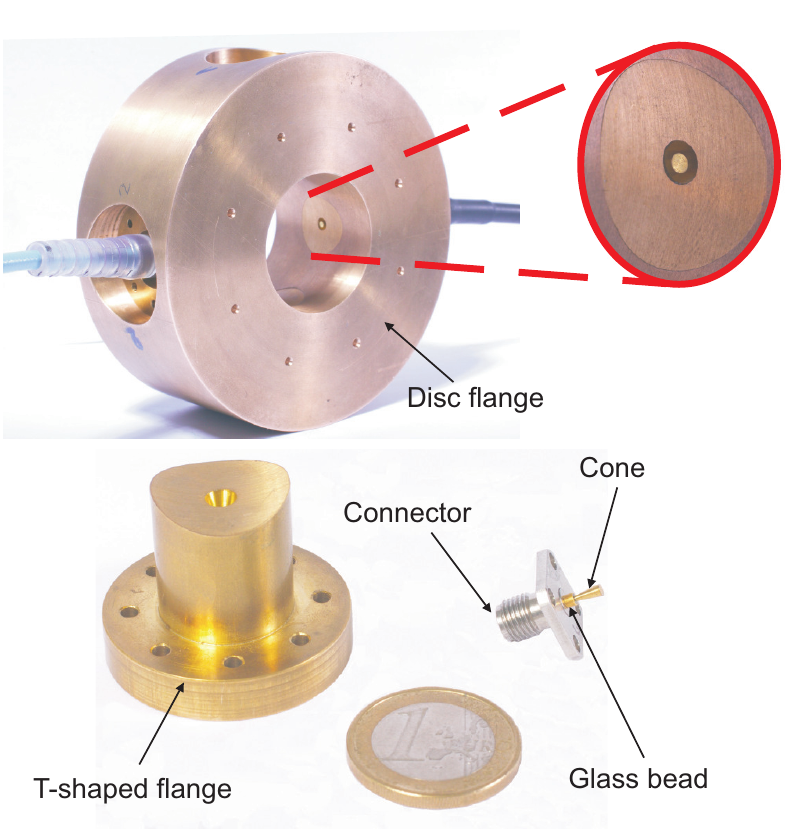}
 \caption{\label{fig:Prototype} Realization of a non-hermetic pickup prototype. Top: Pickup system with four integrated pickups. Bottom: Disassembled pickup element. }
 \end{figure}

In order to validate the simulation results and to investigate the impact of the fabrication tolerances on the rf characteristics, a non-hermetic prototype of the pickup was built (see Fig.~\ref{fig:Prototype}).
The dimensions of the prototype show some deviations from the proposed design. This is due to the limited manufacturing precision of the in-house mechanical workshop. However, the matching conditions are still satisfied because a ratio of $b/a$ of $2.3$ in every cross-section of the pickup was obtained.

The entire structure is made of brass. It consists of a disk flange with four orthogonally distributed T-shaped flanges (see Fig.~\ref{fig:Prototype}). The cone-shaped pickup is soldered to the contact pin of the commercially available glass bead and positioned in the center of the cut-out. The glass bead is designed for $2.92$~mm K-connectors.
The cone-shaped pickups and the connectors are built into the T-shaped flanges, which are mounted in the disk flange. This modular configuration provides for the possibility of changing any of the four pickups in case of damage or upgrade without changing the entire disk flange. This layout is, however, prone to imperfect assembling, which makes a calibration of the pickup necessary.
\begin{figure}[h]
 \centering
 \includegraphics[width=86mm]{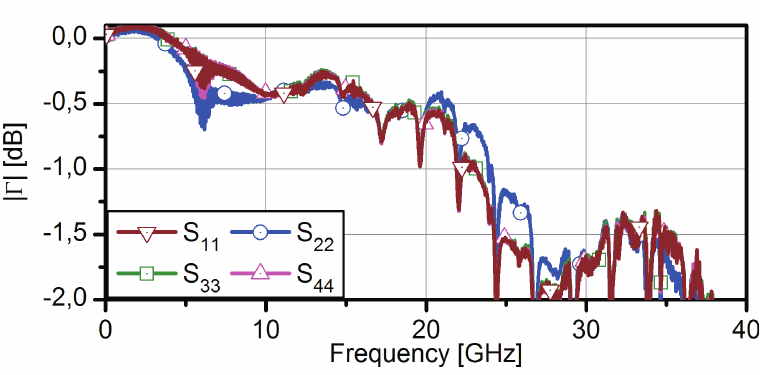}
 \caption{\label{reflection1} Reflection coefficients measured at the connectors. }
 \end{figure}

A series of S-parameter measurements were performed on the pickup system. In the ideal case, the pickup system exhibits a four-fold rotational symmetry. Hence, the rf characteristics should be identical at each port. Figure \ref{reflection1} shows the reflection measured on each of the ports. 
\begin{table}[h]
\caption{\label{MeasDim} Measured cone tip diameter of the fabricated non-hermetic prototype.}
\begin{ruledtabular}
\begin{tabular}{ccccc}
  Port number & 1 & 2 & 3 & 4 \\
\hline
  $a~[mm]$ & $1.87$ & $1.81$ & $1.86$ & $1.87$
\end{tabular}
\end{ruledtabular}
\end{table}
One port is measured at a time, while the others are loaded with $50~\Omega$ terminations. The reflection coefficient curves are in good agreement except for port two (S$_{22}$), which deviates from the other curves in the frequency range from $5$ to $10$~GHz, and from $20$ to $30$~GHz. Reflections higher than $0$~dB for low frequencies occur due to calibration uncertainties.
The cause of the deviation for reflection at port~2 was investigated. It was found that the cone-shaped pickup at port~2 has a smaller tip diameter, $a$. The respective measured sizes are given in Table~\ref{MeasDim}.

In order to investigate the influence on the reflection caused by the different pickup diameter, the mounting position of the cones was swapped and the measurement repeated. A similar result was obtained showing good agreement of three curves and a deviation of the fourth one corresponding to the port, where the defective cone was mounted \cite{Angelovski}. 
\begin{figure}[h]
 \centering
 \includegraphics[width=86mm]{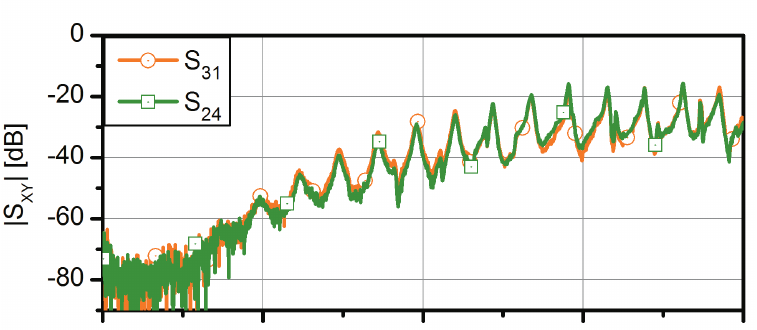}
 \includegraphics[width=86mm]{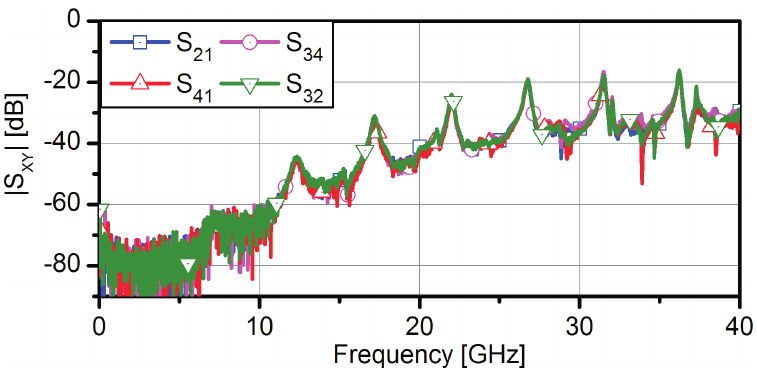}
 \caption{\label{CrossTalk} Measured transmission coefficient between the ports. Top: Between neighboring ports. Bottom: Between opposite ports. }
 \end{figure}

The transmission coefficient between two ports can also be measured by following the same procedure. These measurements show the level of cross-talk between ports. When the transmission between two ports is measured, the other two ports are terminated with $50~\Omega$ impedances. The results are shown in Fig.~\ref{CrossTalk}, where the top panel depicts the cross-talk between two neighboring ports, whereas the bottom panel depicts cross-talk between facing ports. 
\begin{figure}[h]
  \centering
  \includegraphics[width=86mm]{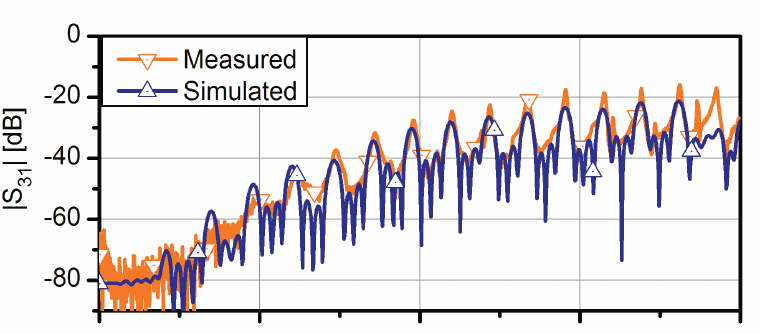}
  \includegraphics[width=86mm]{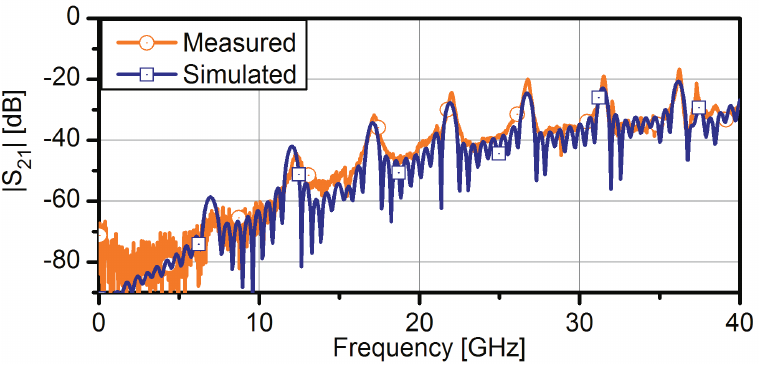}
 \caption{\label{SimComp} Comparison between simulation and measurements. Top: Between neighboring ports. Bottom: Between opposite ports.}
 \end{figure}

The curves are in a good agreement even though the manufactured prototype is not symmetric. From these measurements it is difficult to determine whether there are manufacturing deviations of the system. 
The rf properties in this case are dominated by the dimensions of the aperture of the beam pipe, which behaves like a cavity. The contribution of the pickup, which acts as a coupling probe, is rather small. The maximum level of cross-talk between the ports is $-20$~dB. This is sufficient to attenuate the reflected signal. The resonant peaks at both plots correspond to the one obtained from a resonator with length equal to twice the distance between the ports, respectively. 

For validating the measurement procedure including the defect, the prototype was simulated with CST MICROWAVE STUDIO\textregistered\ software package. Figure \ref{SimComp} shows the comparison between the measurement and the simulation. The results are in a good agreement indicating that the designed pickup can be manufactured and yield the expected performance.

\section{Pickup integration and cabling}
\label{Cabling}

\begin{figure}[h]
 \includegraphics[width=86mm]{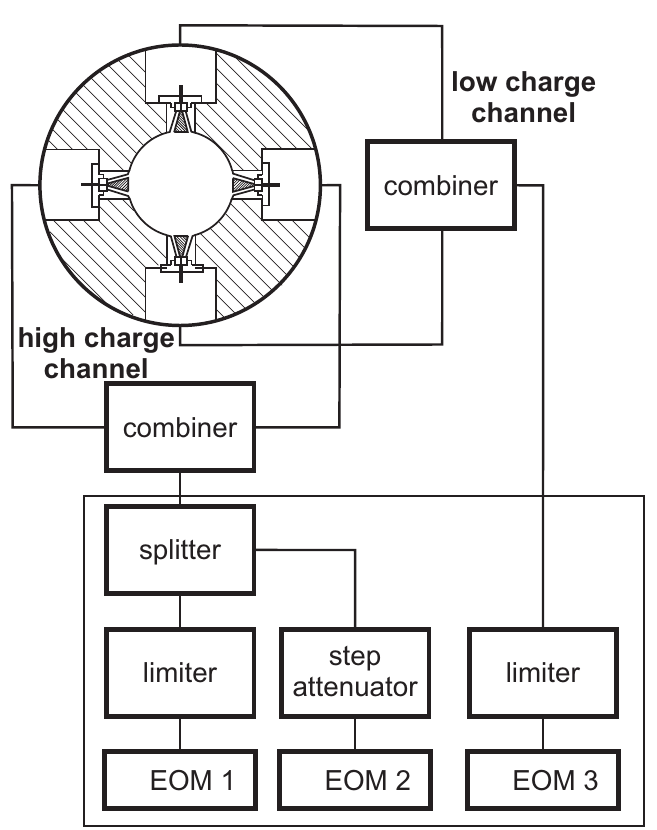}
 \caption{\label{cabling} Proposed rf cabling diagram for low and high charge operation mode of the Bunch Arrival-time Monitor (BAM).}
\end{figure}

The rf front-end comprises cables, combiners, limiters, attenuators and Mach-Zehnder type electro-optic modulators (EOMs). 
In order to realize the required time resolution for low charge operation all rf components of the low charge channel need to have a cutoff frequency similar or higher than the one of the pickup. 
Figure \ref{cabling} shows the rf front-end components and the corresponding cabling diagram for low and high charge operation. 

By combining opposite electrodes of the pickups, the orbit dependency of the arrival time measurement can nearly be eliminated \cite{Kirsten}. 
The voltage signal from the pickup is conducted via phase matched cables ($10^\circ$ at $40$~GHz) and a combiner with a phase unbalance lower than $8^\circ$ up to $40$~GHz. To prevent the EOMs from damage a limiter and a step attenuator will be used.

The cable assembly will be done with silicon dioxide cables which offer high radiation resistance of more than 10000 Gray and a temperature coefficient of delay of about $25$~fs~$m^{-1}K^{-1}$. A slope degradation will occur at the output of the combiner due to a total cable and the combiner phase mismatch of $18^\circ$. The expected slope degradation at the output of the combiner is in the order of $1$\% at a frequency of $40$~GHz. This value is comparable with the ones shown in Table~\ref{fig:toleranzanalyse_slope_und_ringing}.

\section{Summary}

A high bandwidth cone-shaped pickup for the BAMs for free-electron lasers is introduced. The proposed design provides an output voltage slope of $416.78$~mV/ps and a bandwidth of more than $40$~GHz, which is well inside the specifications. This makes it suitable for enabling a sub-$10$~fs time resolution for high and low bunch charge operation of the FELs. A tolerance study was performed providing the sensitivity of the pickup output signal with respect to geometry parameters. This allowed for setting manufacturing tolerances. For validating the obtained results we have built a non-hermetic model of the pickup and conducted a series of measurements. Good agreement between measurements and simulations was found proving the producibility of the proposed pickup system as well as its performance. 

\section{Acknowledgments}

The authors would like to thank the company CST for providing the CST STUDIO SUITE\textregistered\ software package.
The authors would like to thank P.~Gessler, S.~Vilcins-Czvitkovits, M.~Siemens from DESY, Hamburg and J.~R\"onsch-Schulenburg, J.~Ro\ss{}bach of the Institut for Experimental Physics at Hamburg University, Hamburg for support.
The work of A.~Kuhl and  A.~Angelovski is supported by the German Federal Ministry of Education and Research (BMBF) within Joint Project - FSP 301.
Also, A.~Kuhl and S.~M.~Schnepp are supported by the 'Initiative for Excellence' of the German Federal and State Governments and the Graduate School of Computational Engineering at Technische Universit\"at Darmstadt.

\bibliography{}

\end{document}